# Impedance spectroscopy and its application

Pengfei Cheng[1], Shengtao Li[2] and Hui Wang[2]

[1]School of Science, Xi'an Polytechnic University, Xi'an 710048, China
[2]State Key Laboratory of Electrical Insulation and Power Equipment, Xi'an Jiaotong University, Xi'an 710049, China
E-mail: pfcheng@xpu.edu.cn

**Abstract**

Representation of dielectric properties by impedance spectroscopy (IS) is analyzed carefully in this paper. It is found that IS is not a good tool to describe a uniform system because a pseudo relaxation peaks exists at low frequency limit corresponding to direct current (DC) conductivity and two relaxation peaks appears simultaneously corresponding to one relaxation process for a high loss system with $\tan\delta > 1$. However it is very convenient to describe a multiple phase system with IS. When dielectric properties are shown by Cole-Cole equation, only one Cole-Cole arc appears for one phase in IS, therefore it is very easy to distinguish different phases from each other. Especially, since pseudo relaxation exists at low frequency limit for each phase, the location of a certain relaxation process can be deduced by IS without any uncertainty. Furthermore, when dielectric properties are shown with specific impedance spectroscopy (ISI), the information of microstructure can be obtained conveniently. Based on the theoretical results above, dielectric properties of $CaCu_3Ti_4O_{12}$ (CCTO) ceramics with giant dielectric constant (GDC) are investigated. Microstructure of CCTO is obtained by dielectric spectrometer and the origin of GDC is found to come from pseudo relaxation of grain.

**1. Introduction**

Except for macroscopic electrical properties, microscopic mechanisms of polarization and conduction, point defect structure and microstructure of dielectrics can be obtained also by dielectric spectrometer [1-9]. Therefore, it becomes more and more important for today's dielectric physics to realize the fine representation of dielectric properties and to reveal the inherent physical meanings of dielectric representation methods [9].

Generally speaking, there are four representation methods of dielectric properties [10-12]. How to use them to judge a dielectric as a uniform system or a non-uniform one? How to deduce the location of a certain dielectric relaxation? How to distinguish an intrinsic relaxation from Maxwell-Wagner (MW) polarization? These are the most important problems which have to be settled before the analysis of a complex system.

In this paper, impedance spectroscopy (IS) is discussed and compared with dielectric spectroscopy (DS) in detail. It is found that an extra relaxation peak appears at low frequency limit in IS, which is induced by DC conductivity. Furthermore, for high loss dielectric with $\tan\delta > 1$, two relaxation peaks appears which correspond to only one relaxation. If dielectric properties are shown with Cole-Cole equation, only one Cole-Cole arc forms for each phase. Therefore, the number of phases can be obtained easily with IS. As an application, dielectric properties of $CaCu_3Ti_4O_{12}$ ceramics (CCTO) with giant dielectric constant (GDC) are discussed. Although CCTO ceramics have been investigated by many kinds of methods [13-17], the origin of GDC is not clear so far for the complexity of microstructure [18-25].

**2. Theoretical analysis**

*2.1 Impedance spectroscopy or specific impedance spectroscopy for a uniform system*

In order to analyze dielectric properties of dielectric, simple physical model of parallel circuit of resistance and capacitor is often used. According to this model, dielectric properties of a uniform system can be shown in IS as follows

$$Z' = \frac{R}{1+\omega^2 R^2 C^2} = \frac{R}{1+\omega^2 \tau_P^2} \quad (1)$$

$$Z'' = R\frac{\omega RC}{1+\omega^2 R^2 C^2} = R\frac{\omega \tau_P}{1+\omega^2 \tau_P^2} \quad (2)$$

where $R$ and $C$ are the resistance and capacitor of the dielectric respectively, $\omega$ is angle frequency,

$\tau_p$=RC is circuit responding time. In consideration of relaxation conductivity $g$ and DC conductivity $\gamma$ at the same time, $\tau_p$ can be expressed as

$$\tau_P = RC = \rho \frac{l}{S} \frac{\varepsilon_0 \varepsilon' S}{l} = \rho \varepsilon_0 \varepsilon' = \frac{\varepsilon_0 \varepsilon'}{\gamma + g} = \frac{\varepsilon_0 \varepsilon'}{\omega \varepsilon_0 \varepsilon''} = \frac{1}{\omega \tan \delta} \quad (3)$$

where $\varepsilon_0$ is dielectric constant of free space, $\varepsilon'$ and $\varepsilon''$ are real and imaginary parts of dielectric constant respectively, $\rho$ is resistivity, $l$ is the thickness, $S$ is electrode area, and $\tan\delta$ is loss tangent. It is clear from Eqn. (2) that a relaxation peak will appear in $Z''$-$f$ curve when $\omega\tau_p$=1. This condition is equivalent to that $\tan\delta$=1. For a dielectric relaxation with $(\tan\delta)_m$>1, there will be two frequency points corresponding to $\tan\delta$=1. In this case, two relaxation peaks will be observed in $Z''$-$f$ curve which correspond to only one relaxation process. At low frequency limit, DC conductance is the main process and dielectric relaxation can be ignored, then IS can be shown as

$$Z' = \frac{R_0}{1+\omega^2 R_0^2 C_0^2} = \frac{R_0}{1+\omega^2 \tau_{P0}^2} \quad (4)$$

$$Z'' = R_0 \frac{\omega R_0 C_0}{1+\omega^2 R_0^2 C_0^2} = R_0 \frac{\omega \tau_{P0}}{1+\omega^2 \tau_{P0}^2} \quad (5)$$

where $R_0$ and $C_0$ are DC resistance and geometric capacitance respectively and $\tau_{p0}$ is

$$\tau_{p0} = R_0 C_0 = \frac{\varepsilon_0 \varepsilon_s}{\gamma} \quad (6)$$

It can be known from Eqn. (5) that a peak will appear in $Z''$-$f$ curve at low frequency limit when $\omega\tau_{p0}$=1. This implies that DC conductivity can induce a peak in IS, which is called pseudo relaxation peak in this paper. Therefore, for a uniform system with $n$ kinds of dielectric relaxations, $2n$+1 relaxation peaks may be obtained at most by IS.

What is the relationship between circuit responding time $\tau_p$ and dielectric relaxation time $\tau$? If DC conductivity can be ignored, we have

$$\omega\tau_P = \frac{\varepsilon_s + \varepsilon_\infty \omega^2 \tau^2}{(\varepsilon_s - \varepsilon_\infty)\omega\tau} \quad (7)$$

According to Eqn. (7), a relaxation peak in $Z''$-$f$ curve will be observed when the following condition is satisfied

$$\frac{\varepsilon_s + \varepsilon_\infty \omega^2 \tau^2}{(\varepsilon_s - \varepsilon_\infty)\omega\tau} = 1 \quad (8)$$

The premise of real solution of the equation above is that $k=\varepsilon_s/\varepsilon_\infty > 3+2\sqrt{2}$. Under this condition, we have $\omega\tau=[(k-1)\pm\sqrt{k^2-6k+1}]/2 > 1$. This implies that relaxation peak in IS moves to high frequency compared with that in DS. However, no relaxation peak will be observed in IS if $k=\varepsilon_s/\varepsilon_\infty < 3+2\sqrt{2}$.

IS can be shown in the form of Cole-Cole equation as follows

$$Z''^2 + \left[Z' - \frac{1}{2(G_0+G_r)}\right]^2 = \left[\frac{1}{2(G_0+G_r)}\right]^2 \quad (9)$$

where $G_0$ and $G_r$ are DC conductance and relaxation conductance respectively. When no dielectric relaxation exists the equation above represents a standardized circular arc with radium of $R_0/2$ and the center at $(R_0/2, 0)$. However, if one or more dielectric relaxations exist in a uniform system, the Cole-Cole arc is no longer a circular one for the variation of effective conductance with frequency. The shape of Cole-Cole arc varies with frequency gradually from the circle defined by DC resistance $R_0$ to the circle defined by dielectric relaxation process. Whether the arc is circular or not, the intersection points at $Z'$ axis are $(0, 0)$ and $(R_0, 0)$. From this information, DC conduction process can be learned. Based on the analysis above, it can be concluded that only one Cole-Cole arc can be formed no matter how many dielectric relaxations exist in the uniform system. For a non-uniform system the number of phases can be deduced from the number of Cole-Cole arcs.

In fact, for a uniform system with dielectric relaxations, IS as shown in Eqn. (1) and (2) is not a convenient tool for the variation of $R$ and $C$ with frequency. In order to learn the intrinsic characteristics of dielectric, the influence of physical dimension has to be removed. After the elimination of physical dimension from IS, specific impedance spectroscopy (SIS) can be expressed as follows

$$z' = \frac{1}{\omega\varepsilon_0}\frac{\varepsilon''}{\varepsilon'^2+\varepsilon''^2} \quad (10)$$

$$z'' = \frac{1}{\omega\varepsilon_0}\frac{\varepsilon'}{\varepsilon'^2+\varepsilon''^2} \quad (11)$$

At low frequency limit, the influence of dielectric relaxation can be ignored and SIS can be expressed as

$$z' = \frac{1}{\gamma(1+\omega^2\tau_z^2)} \quad (12)$$

$$z'' = \frac{\omega\tau_z}{\gamma(1+\omega^2\tau_z^2)} \quad (13)$$

where $\tau_z = \varepsilon_0\varepsilon_s/\gamma$. The equations above are similar to Debye equations. The slope of $\ln z''$-$\ln f$ curve is +1 when $\omega \to 0$, while the slope is -1 when $\omega \to \infty$, therefore a obvious loss peak appears when $\omega\tau_z=1$. Pseudo relaxation peak is observed clearly again in SIS, as shown in Fig. 1(a). If DC conductivity is ignored at high frequency region, dielectric relaxation can be described as follows

$$z' = \frac{(\varepsilon_s-\varepsilon_\infty)\tau}{\varepsilon_0}\frac{1}{\varepsilon_s^2+\varepsilon_\infty^2\omega^2\tau^2} \quad (14)$$

$$z'' = \frac{\varepsilon_s}{\omega\varepsilon_0(\varepsilon_s^2+\varepsilon_\infty^2\omega^2\tau_1^2)} + \frac{\varepsilon_\infty\tau^2}{\varepsilon_0}\frac{\omega}{\varepsilon_s^2+\varepsilon_\infty^2\omega^2\tau^2} = z_1'' + z_2'' \quad (15)$$

where $z_1''$ is a decreasing function. Let $dz_2''/d\omega = 0$, we have $(z_2'')_{max} = \varepsilon_\infty\tau/2\varepsilon_0\varepsilon_s^2$ when $\omega\tau = \varepsilon_s/\varepsilon_\infty$. This indicates that relaxation peaks in SIS moves to higher frequency compared with DS. When $\omega \to 0$, $\ln z'' \approx \ln z_1''$, and the slope of $\ln z''$-$\ln f$ curve is -1. When $\omega \to \infty$, $\ln z'' \approx \ln z_2''$, and the slope of $\ln z''$-$\ln f$ curve is -1 too. Therefore, $\ln z''$-$\ln f$ curve corresponding to a dielectric relaxation is a transition from a line with slope of -1 to another line with the same slope, as shown in Fig. 1(b). The curve of $\ln z''$-$\ln f$ is slanting Z-shaped and no apparent relaxation peak appears.

DC conduction process is expressed with a pseudo relaxation peak at low frequency limit, while dielectric relaxation process is expressed with two parallel lines with slope of -1. The difference in curve shape gives a convenient way to distinguish them from each other clearly, as shown in Fig. 1.

*2.2. Specific impedance spectroscopy for a non-uniform system*

For a non-uniform system, total impedance $Z$ can be expressed as the sum of each phase as follows

$$Z' = \frac{R_1}{1+\omega^2 R_1^2 C_1^2} + \frac{R_2}{1+\omega^2 R_2^2 C_2^2} \quad (16)$$

$$Z'' = R_1 \frac{\omega R_1 C_1}{1+\omega^2 R_1^2 C_1^2} + R_2 \frac{\omega R_2 C_2}{1+\omega^2 R_2^2 C_2^2} \quad (17)$$

If the thickness of the $i$th phase is denoted as $l_i$, and the ratio of $l_i$ to the total thickness $l$ of the system is denoted as $k_i$, then we have

$$z' = \frac{1}{\omega\varepsilon_0}\left(\frac{k_1 \varepsilon_1''}{\varepsilon_1'^2 + \varepsilon_1''^2} + \frac{k_2 \varepsilon_2''}{\varepsilon_2'^2 + \varepsilon_2''^2}\right) = k_1 z_1' + k_2 z_2' \quad (18)$$

$$z'' = \frac{1}{\omega\varepsilon_0}\left(\frac{k_1 \varepsilon_1'}{\varepsilon_1'^2 + \varepsilon_1''^2} + \frac{k_2 \varepsilon_2'}{\varepsilon_2'^2 + \varepsilon_2''^2}\right) = k_1 z_1'' + k_2 z_2'' \quad (19)$$

where $z_i'$ and $z_i''$ are real and imaginary parts of $i$th phase specific impedance, and $\varepsilon_i'$ and $\varepsilon_i''$ are real and imaginary parts of $i$th phase dielectric constant. It can be seen clearly from eqn. (16) to eqn. (19) that the dielectric properties of each phase can be shown one by one along frequency axis for a non-uniform system. Therefore, IS or SIS are the most convenient methods to represent the dielectric characteristics of a non-uniform system. After estimation of the thickness of each phase according to eqn. (18) and (19), the microstructure of the non-uniform system can be learned.

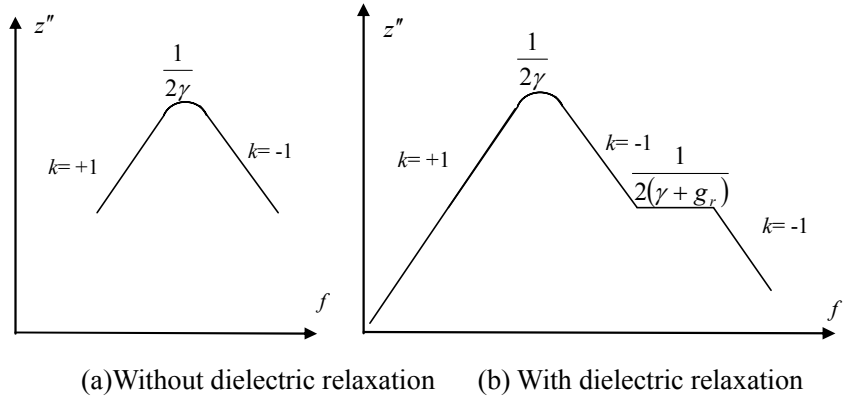

(a) Without dielectric relaxation    (b) With dielectric relaxation

Fig. 1 Specific impedance spectroscopy of a uniform system

The influence of DC conductance for a non-uniform system in SIS is shown in Fig. 2 (a). Except for the low frequency pseudo relaxation induced by DC conductance of the first phase, another pseudo relaxation is observed which is induced by DC conductance of the second phase. If the SIS for a non-uniform system is regarded as that of a uniform dielectric, the second pseudo relaxation has to be regarded as an apparent relaxation, namely Maxwell-Wagner (MW) polarization. In other words, MW polarization originates form pseudo relaxation. For an electronic ceramics, the basic phases are grainboundary and grain. Since DC conductivity of grainboundary is much lower than grain, SIS curves of grainboundary must locates at lower frequency compared with that of grain. If a dielectric relaxation exists in grainboundary, it must be at higher frequency than pseudo relaxation of grainboundary, while at lower frequency than pseudo relaxation of grain, as shown in Fig. 2(b). Of course, the relaxation in grain can be represented also in the same way. Thus the location of a certain relaxation can be deduced without any uncertainty by SIS.

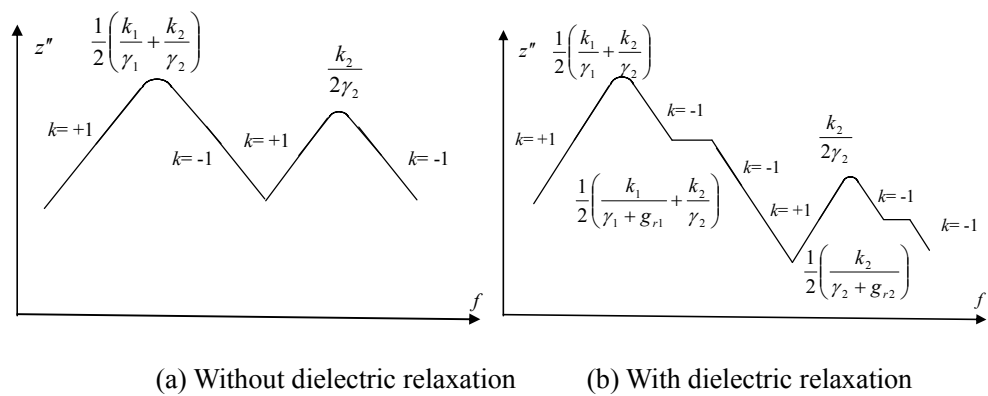

(a) Without dielectric relaxation  (b) With dielectric relaxation

Fig. 2 Specific impedance spectroscopy of a non-uniform system

## 4. Dielectric properties of $CaCu_3Ti_4O_{12}$ ceramics

CCTO ceramics are prepared by traditional solid phase sintering methods and the dielectric properties are measured with Novocontrol wide band dielectric spectrometer. Measuring frequencies vary from 0.1 Hz to $10^7$ Hz and measuring temperatures change from -100°C to 100°C. Dielectric properties of CCTO ceramics are shown in Fig. 3 (a) and (b). Dielectric constant of CCTO ceramics at low frequency is more than $10^5$, while it is just about 100 at high frequency. Two relaxation peaks can be observed from $\varepsilon''$-$f$ curve, which means that two dielectric relaxations exist in CCTO ceramics. At the same time, two Cole-Cole arcs appear in $z''$-$z'$ curve, as shown in Fig.3 (e) and (f), implying that two phases coexist in CCTO ceramics. In $z''$-$f$ curve, an obvious peak can be observed at low frequency, as shown in Fig. 3 (d). The slope is 0.90 for the increasing region and -0.96 for the decreasing region, both of which are very close to the theoretical values of +1 and -1 of pseudo relaxation, therefore it can be concluded that the low frequency relaxation peaks originates form DC conductance of CCTO ceramics. It can be calculated form Fig.2 (c) that the activation energy of DC conductance of CCTO ceramics is 0.68eV, which corresponds to barrier height at grainboundary. In fact, another relaxation peak can be obtained also at high frequency, and the corresponding activation energy is 0.1eV. According to the same reasoning process, it is another pseudo relaxation induced from DC conductance of grain. Obviously, pseudo relaxation at low frequency corresponds to grainboundary, while pseudo relaxation at high frequency corresponds to grain. Compared with the loss peak in DS, relaxation peak moves to high frequency. For example, loss peak locates at $4.51\times10^3$Hz in DS when measured at 100°C, while the corresponding relaxation peak locates in SIS is at 4.44×10$^6$Hz. Consideration the ratio of $\varepsilon_s/\varepsilon_\infty$ is about 1000, it can be deduced that the pseudo relaxation at high frequency in SIS corresponds to the dielectric relaxation at high frequency in DS. In other words, giant dielectric constant of CCTO ceramics comes from internal boundary layer capacitor (IBLC) effect of MW polarization. Low frequency relaxation, as shown in Fig.3 (c) is between pseudo relaxations of grainboundary and grain, so it must locates at grainboundary.

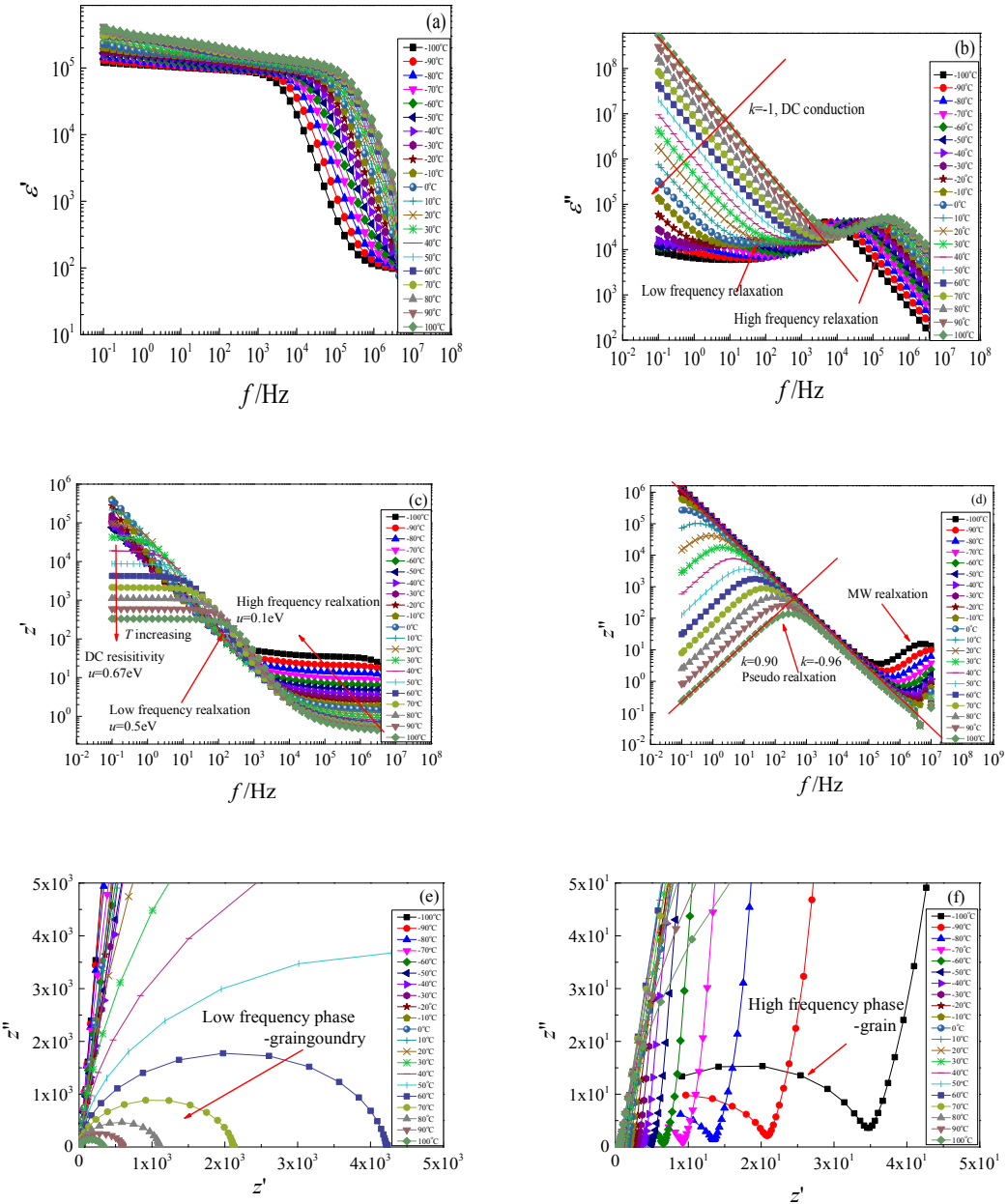

Fig.3 Dielectric properties of CaCu$_3$Ti$_4$O$_{12}$ ceramics. (a) $\varepsilon'$-$f$. (b) $\varepsilon''$-$f$. (c) $z'$-$f$. (d) $z''$-$f$. (e) Low frequency part of $z''$-$z'$. (f) High frequency part of $z''$-$z'$.

There are two platforms in $z'$-$f$ curve. According to Eqn. (11), the first platform at low frequency corresponds to $k_1/\gamma_1$, and the second platform at high frequency corresponds to $k_2/\gamma_2$. Dielectric constants at low frequency and high frequency obtained from $\varepsilon'$-$f$ curve at -100$^\circ$C are 94 and 121430 respectively, then $k_1$ can be estimated to be $7.74\times10^{-4}$ according IBLC effect. The grain size of CCTO ceramics is about 100μm [26], so grainboundary thickness can be calculated to be 77nm. DC resistivity of grain and grainboundary at room temperature can be estimated also. From $z'$-$f$ curve at 30$^\circ$C, it can be read that $k_2z_2\approx z_2$= 0.99221 Ω·m and $k_1z_1+ k_2z_2\approx k_1z_1$=41849, so the resistivity of grainboundary can be obtained to be $z_1$=41849/$k_1$=5.406$\times10^7$ Ω·m, which is about $10^7$ times larger than grain.

## 4. Conclusions

Impedance spectroscopy (IS) is analyzed in detail in this paper. It is found that pseudo relaxation may be induced at low frequency limit by DC conductance. According to the difference in shape of IS for pseudo relaxation and dielectric relaxation, DC transport process can be distinguished from dielectric relaxation. Since $2n+1$ relaxation peaks may be obtained at most by IS for a uniform system with $n$ kinds of dielectric relaxations, IS is not a good tool to represent dielectric properties of a uniform system. However IS is very convenient for a non-uniform system because only one Cole-Cole curve exists for each phase, no matter how many dielectric relaxations exist in the phase. At the end of this paper, the dielectric properties of $CaCu_3Ti_4O_{12}$ (CCTO) ceramics are discussed. It is found that giant dielectric constant (GDC) originates form pseudo relaxation of grain. At the same time, grainboundary thickness and resistivity of grainboundary and grain are calculated theoretically. The results of this paper provide a solid theory foundation for analysis of a complex dielectric.

**Acknowledgements**

This work is supported by the National Natural Science Foundation of China (No. 51277138, 50972118), the Scientific Research Plan Projects of Education Department of Shaanxi Province of China (No. 12JK0434, 2010JK573) and the Doctoral Scientific Research Foundation of Xi'an Polytechnic University (No. BS0814).

Fig. 1 Specific impedance spectroscopy of a uniform system. (a) Without dielectric relaxation (b) With dielectric relaxation

Fig. 2 Specific impedance spectroscopy of a non-uniform system. (a) Without dielectric relaxation (b) With dielectric relaxation

Fig.3 Dielectric properties of $CaCu_3Ti_4O_{12}$ ceramics. (a) $\varepsilon'$-$f$. (b) $\varepsilon''$-$f$. (c) $z'$-$f$. (d) $z''$-$f$. (e) Low frequency part of $z''$-$z'$. (f) High frequency part of $z''$-$z'$.

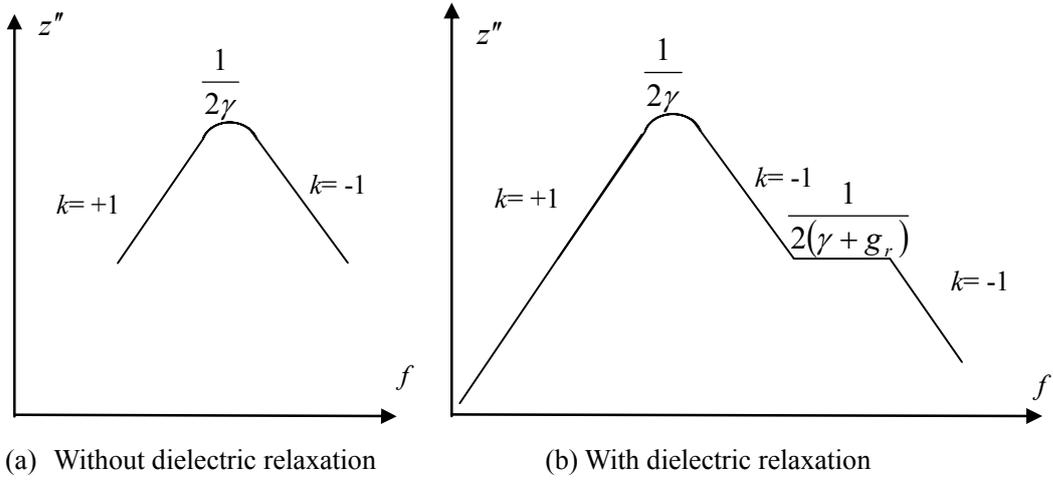

(a) Without dielectric relaxation    (b) With dielectric relaxation

Fig. 1 Specific impedance spectroscopy of a uniform system

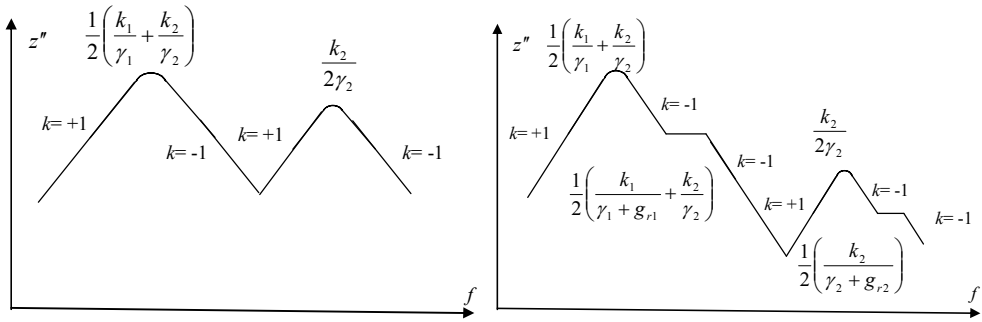

(a) Without dielectric relaxation      (b) With dielectric relaxation

Fig. 2 Specific impedance spectroscopy of a non-uniform system

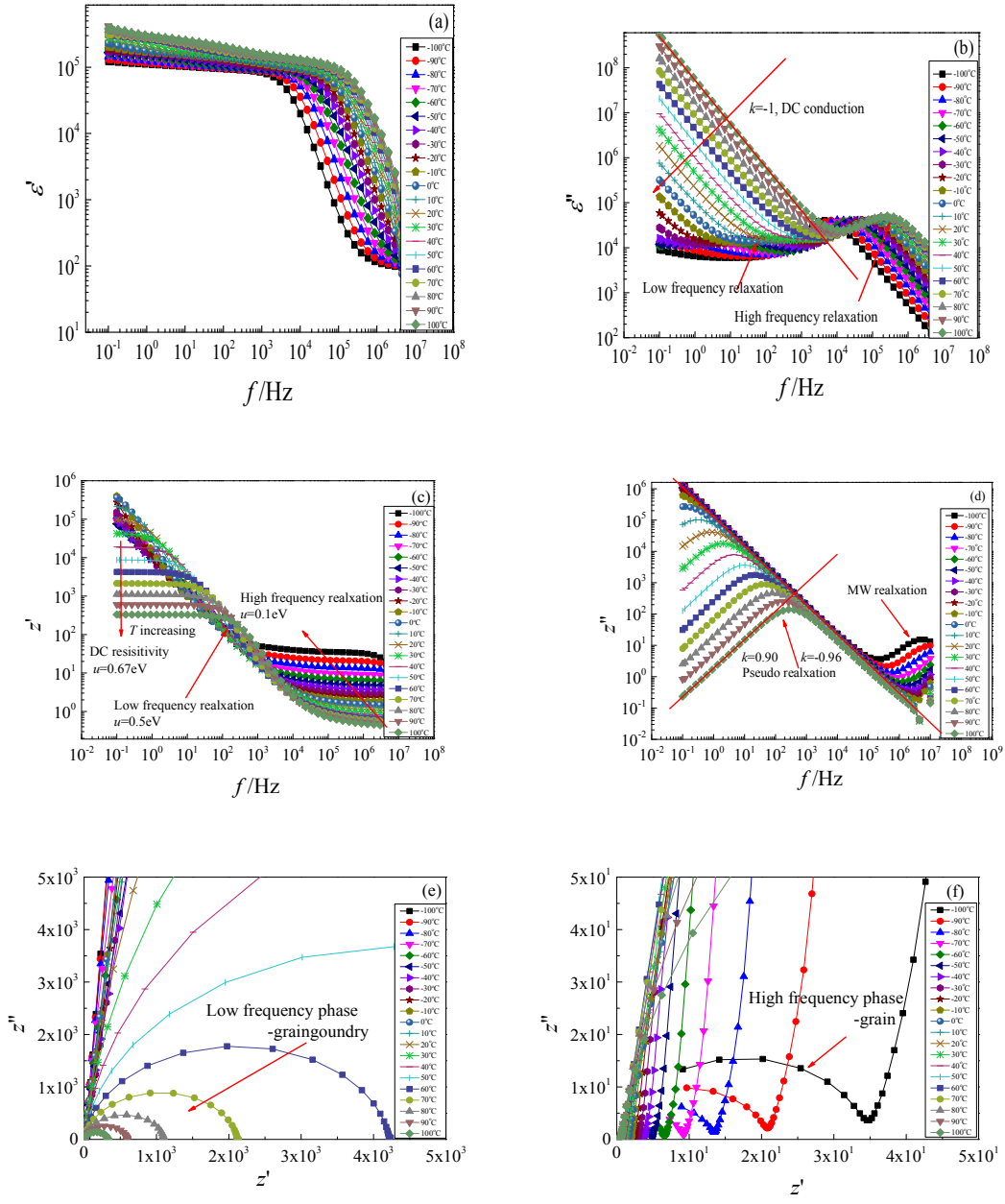

Fig.3 Dielectric properties of CaCu$_3$Ti$_4$O$_{12}$ ceramics. (a) $\varepsilon'$-$f$. (b) $\varepsilon''$-$f$. (c) $z'$-$f$. (d) $z''$-$f$. (e) Low frequency part of $z''$-$z'$. (f) High frequency part of $z''$-$z'$.